\documentclass[%
 reprint,
 amsmath,amssymb,
 pre,
]{revtex4-1}

\usepackage{graphicx}
\usepackage{dcolumn}
\usepackage{bm}
\usepackage{color}

\begin{document}

\title{Effect of particle size distribution on polydisperse hard disks}

\author{Pablo Sampedro Ruiz}
\affiliation{School of Chemical and Biomedical Engineering, Nanyang Technological University, \\62 Nanyang Drive, 637459, Singapore 
}

\author{Ran Ni}
\email{r.ni@ntu.edu.sg}
\affiliation{School of Chemical and Biomedical Engineering, Nanyang Technological University, \\62 Nanyang Drive, 637459, Singapore 
}

\begin{abstract}
Using Monte Carlo simulations, we systematically investigate the effect of particle size distribution on the phase behaviour of polydisperse hard disks.
Compared with the commonly used Gaussian-like polydisperse hard disks [Commun. Phys. 2, 70 (2019)], we find that the phase behaviour of polydisperse hard-disk systems with lognormal and triangle distributions are significantly different. In polydisperse hard-disk systems of lognormal distributions, although the phase diagram appears similar to that of Gaussian-like polydisperse hard disks, the re-entrant melting of hexatic or solid phase can not be observed in sedimentation experiments. For polydisperse hard-disk systems of triangle distributions, the phase behaviour is qualitatively different from the Gaussian-like and lognormal distributions, and we can not reach any system of true polydispersity larger than 0.06, which is due to the special shape of the triangle distribution.
Our results suggest that the exact particle size distribution is of primary importance in determining the phase behaviour of polydisperse hard disks, and we do not have a universal phase diagram for different polydisperse hard-disk systems.
\end{abstract}

\pacs{Valid PACS appear here}
\maketitle


\section{Introduction}

Since the proposal of the celebrated Kosterlitz-
Thouless-Halperin-Nelson-Young (KTHNY) theory~\cite{kt,hn2,hn,young,bkt}, much attention has been focused in the study of 2D solids, due to the unique physics of the system~\cite{Strandburg1988,Dash1999,Grasser2009}. Namely, the theory predicts the existence of a new phase of matter, i.e., hexatic phase, that intersects with both the solid and fluid, with the system evolving through two continuous transitions. { However, it has been found that the actual behaviour depending heavily on the nature of interaction between particles, and various theories have been proposed to explain different scenarios~\cite{chiu1982,saito1982,binder2002}.}

In the system of monodisperse hard disks, arguably one of the `simplest' particle systems for studying phase transitions in 2D, there have been a long debate about the mechanism and nature of the melting transition~\cite{zahn1999,karn2000,han2008,RICE20091,murray1987,marcus1996,maret2004,keim2007,stuart2008}, which was recently settled that it occurs in two steps with a continuous solid-hexatic transitions closely followed by a first order transition to the fluid\cite{hdprl,hdpre}, and the shape and softness of particles also play important roles in the 2D melting~\cite{krauth2015,glotzer2017,massimo2018}. 
Besides, simulations of binary hard-disk mixtures showed that the presence of tiny amounts of small particles can eliminate the hexatic phase~\cite{russo2017}.
However, in our recent work, we found that in the continuous polydisperse system of hard disks, the particle size polydisperse can change the nature of melting transition. With increasing the degree of polydispersity, the first-order hexatic-fluid transition becomes weaker and eventually disappears at relatively high polydispersity, and simultaneously the density range of stable hexatic phase increases by orders of magnitude~\cite{SampedroRuiz2019}.
More intriguingly, for systems of relatively highly polydispersed hard disks, a re-entrant melting transition was observed with increasing the density of system, which was proven impossible in 3D system of polydisperse hard spheres~\cite{bartlett1999,sollich2003prl}.

In Ref.~\cite{SampedroRuiz2019}, we assumed that the polydisperse hard-disk system is in contact with a dilute reservoir of Gaussian-like particle size distribution.
To model the effect of polydispersity, we considered a 2D system of volume $V$ containing $N$ polydisperse hard disks based on the semigrand canonical ensemble, in which the chemical potential difference between particles of different size is fixed~\cite{kofke1988,bolhuis1996,kofke1999,frenkel2004}, and its distribution obeys
\begin{equation}
\frac{\Delta \mu (\sigma)}{k_BT} = -\frac{\left(\sigma-\sigma_0\right)^2}{2 \nu^2}
\label{eq:chapt4_1}
\end{equation}
where $\sigma$ is the particle diameter changing from 0 to $\infty$, with $k_B$ and $T$ the Boltzmann constant and temperature of the system, respectively. $\nu$ is the polydispersity parameter and the main controlling parameter of the system.
In the ideal gas limit Eq.~\ref{eq:chapt4_1} gives a Gaussian-like particle size distribution centered around $\sigma_0$ with the standard deviation $\nu$.
However, the artefact of Eq.~\ref{eq:chapt4_1} is that there is a finite probability of having $\sigma = 0$, and it is unrealistic and pronounced at large $\nu$, where the re-entrant melting was observed.
This opens the question whether the reported re-entrant melting generally exists in polydisperse hard-disks systems or only exists in the system of artificial Gaussian-like polydisperse hard disks. To this end, we investigate the effect of particle size distribution on the phase behavior of polydisperse hard disks, and we simulate polydisperse hard-disk systems of two representative particle size distribution, i.e., the lognormal and triangle distributions, in which the probability of having $\sigma = 0$ is exactly zero. The difference is that in the triangle distribution, there is a finite range of $\sigma$ having probability zero, while in the lognormal distribution, the particle size probability only vanishes at $\sigma = 0$.

The article is organized as follows. In Sec.~\ref{sec1}, we summarize the phase behavior of the Gaussian-like (Eq.~\ref{eq:chapt4_1}) polydisperse hard disks, which was reported in Ref~\cite{SampedroRuiz2019}, after which we investigate the phase behavior of two other representative polydisperse hard-disk systems, i.e., lognormal distribution in Sec.~\ref{sec2}, and triangle distribution in Sec.~\ref{sec3}. Conclusions are drawn in Sec.~\ref{sec4}.

\section{Gaussian-like polydisperse hard disks \label{sec1}}
\begin{figure}
\begin{center}
\includegraphics[width=1.\linewidth]{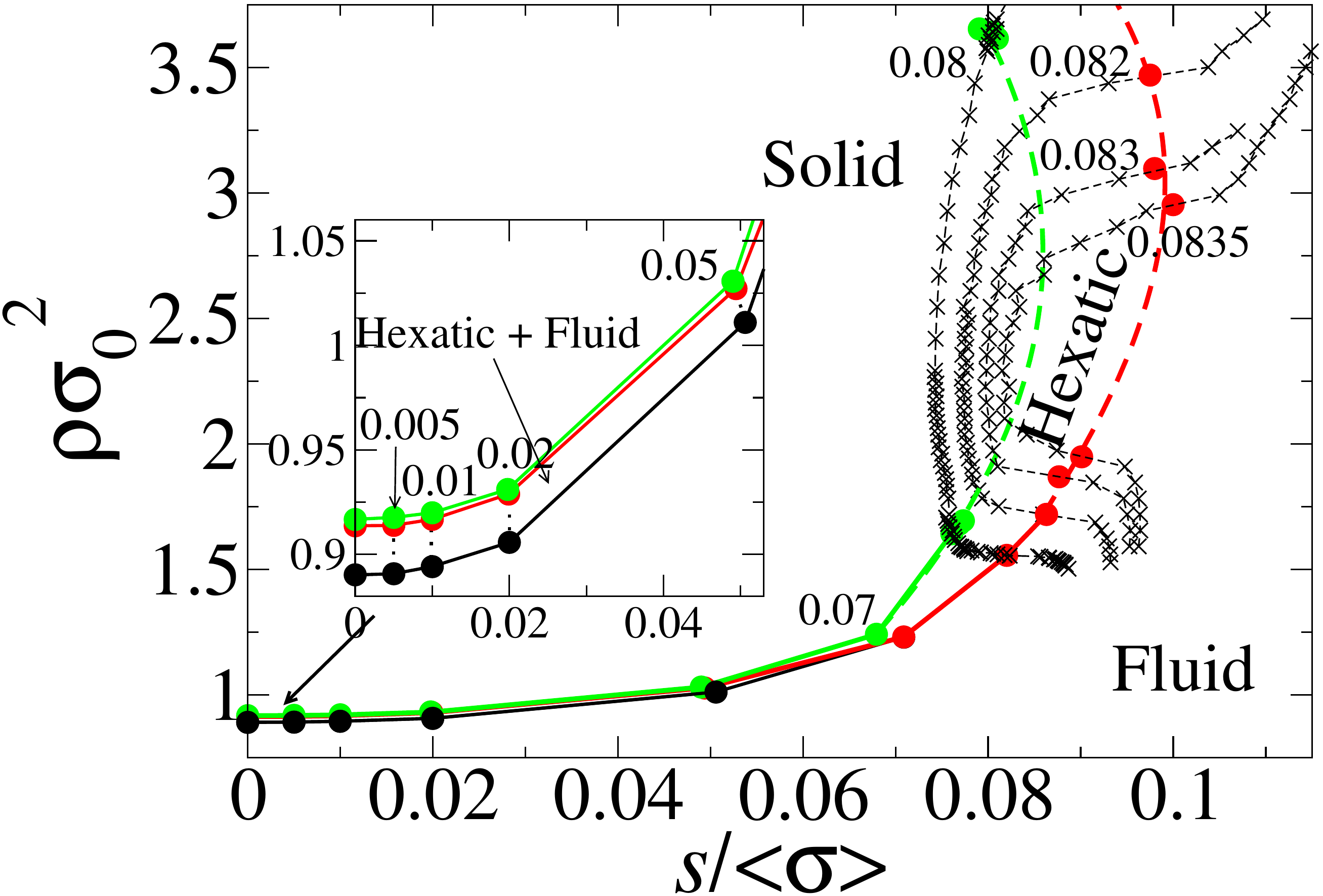} 
\caption{Phase diagram of the Gaussian-like polydisperse hard-disk systems in the representation of $\rho \sigma_0^2$ and $s/\langle \sigma \rangle$, in which the dashed green and red lines are the interpolated phase boundaries for re-entrant melting of solid and hexatic phases, respectively. The phase boundaries are obtained from $NVT-\Delta \mu$ simulations for systems with $\nu/ \sigma_0 = 0.005$  to 0.0835. Inset: the enlarged view of the region of phase diagram at $0 \le s/\langle \sigma \rangle \le 0.02$. The crosses are the simulation results 
along the the re-entrant transitions at $\nu/\sigma_0 = 0.08$, 
0.0805, 0.081, 0.082, 0.083, and 0.0835 from left to right~\cite{SampedroRuiz2019}.
\label{fig:chapt4_1}
}
\end{center}
\end{figure}

In Ref.~\cite{SampedroRuiz2019}, we investigated the phase behavior of Gaussian-like polydisperse hard-disk systems, in which we essentially considered semigrand canonical systems of volume $V$ consisting of $N$ polydisperse hard disks, and the chemical potential difference between particles of different size is controlled by Eq~\ref{eq:chapt4_1}. 
All the simulations in this article are done for systems with $N=256^2$ particles, and to obtain the phase diagram of polydisperse hard disks, we perform $NVT-\Delta \mu$ simulations with the event chain Monte Carlo (ECMC) algorithm~\cite{SampedroRuiz2019,manon2013}, { where $\Delta \mu$ is the chemical potential difference between particles of different size.}
The phase diagram of the system in the representation of $\rho \sigma_0^2$ and $s/\langle \sigma \rangle$ is summarized in Fig.~\ref{eq:chapt4_1}, where $\rho = N/V$ is the density of the system, and $s = \sqrt{\langle \sigma^2 \rangle - \langle \sigma \rangle^2}$ is the size polydispersity of the system. The two major features of Fig.~\ref{fig:chapt4_1} are: (1) with increasing $\nu$, the two-stage melting with decreasing density as found in the system of monodisperse hard disks, i.e., a continuous solid-hexatic transition followed by a first-order hexatic-fluid transition, gradually changes to be celebrated KTHNY scenario consisting of a continuous solid-hexatic transition followed by a continuous hexatic-fluid transition with a significantly enlarged density range of stable hexatic phase; (2) in systems of highly polydisperse hard disks, e.g. $\nu/\sigma_0$ around 0.08, with increasing density, the system undergoes a re-entrant melting transition by forming two fluid phases at both low and high densities.

\begin{figure}
\begin{center}
\includegraphics[width=1.\linewidth]{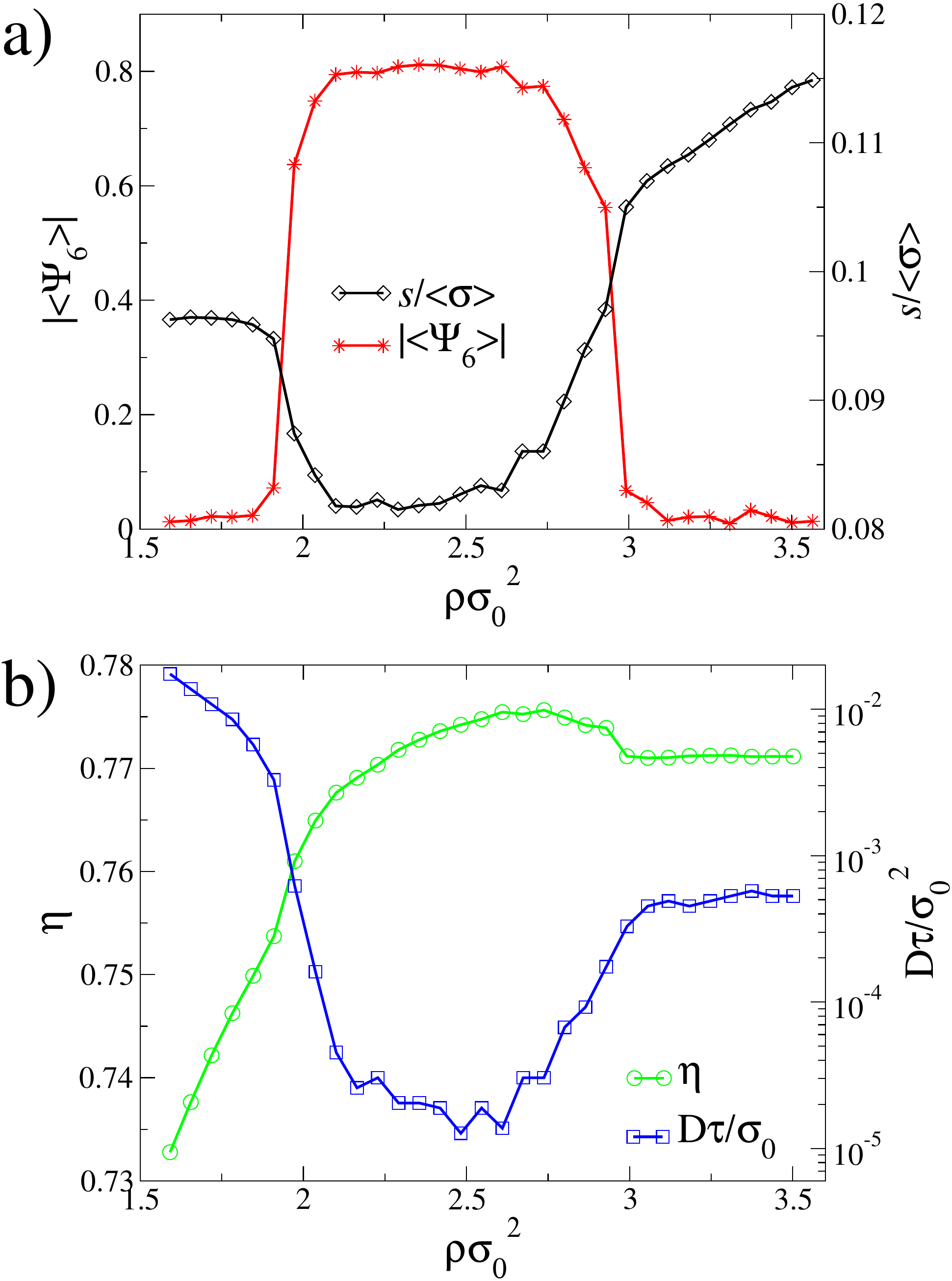} 
\caption{$|\langle \Psi_6 \rangle|$ and $s/\langle \sigma \rangle$ (a) as well as $\eta$ and $D\tau/\sigma_0^2$ (b) as functions of density $\rho \sigma_0^2$ for Gaussian-like polydisperse hard disks with $\nu/\sigma_0=0.0835$. Here $D$ is the diffusion coefficient of the system with $\tau$ the time unit of event driven molecular dynamics simulations.
\label{fig2}
}
\end{center}
\end{figure}

The most intriguing result is the re-entrant transition in highly polydisperse systems of hard disks, which has been proven impossible in 3D systems of polydisperse hard spheres~\cite{bartlett1999,sollich2003prl}. To show the structural change of the highly polydisperse system ($\nu/\sigma_0=0.0835$) with density, we plot the sixfold bond orientation order parameter $\langle \Psi_6 \rangle$ as a function of density $
\rho \sigma_0^2$ in Fig.~\ref{fig2}a, and 
\begin{equation}
\langle \Psi_6 \rangle = \left \langle \frac{1}{N} \sum_{k=1}^N \psi_6 (\mathbf{r}_k) \right  \rangle,
\end{equation}
with $\psi_6 (\mathbf{r}_k) = \frac{1}{N_k} \exp (i 6 \theta_{kj})$, where $\theta_{kj}$ is the angle between the vector connecting particle $k$ with its neighbor $j$ and a chosen fixed reference vector, and $N_k$ is the number of first neighbours for particle $k$ based on the Voronoi tessellation of the system. One can see that with increasing density, the system transforms from a disordered fluid to an ordered phase and then into an disordered phase again at very high density. Here the ordered phase is hexatic phase, but it is not determined by 
 $\langle \Psi_6 \rangle$, which is essentially an indication of the formation of an ordered phase. The quantitative determination on the the phase of solid, hexatic  or fluid is done by checking the positional correlation $g(x,0)-1$, which exhibits a power law decay $\sim x^{-\alpha}$ with $\alpha \le 1/3$ in the solid phase { in the KTHNY scenario~\cite{kt,hn2,hn,young,bkt}} and an exponential decay $\sim \exp(-x)$ in a hexatic or fluid phase, and the sixfold orientation correlation function { $g_6(r) = \langle \psi_6^*(\mathbf{r}' + \mathbf{r}) \psi_6(\mathbf{r}')\rangle$}, which exhibits a power law decay in a hexatic phase and an an exponential decay in a fluid~\cite{SampedroRuiz2019}. 
 
\begin{figure}
\begin{center}
\includegraphics[width=1.\linewidth]{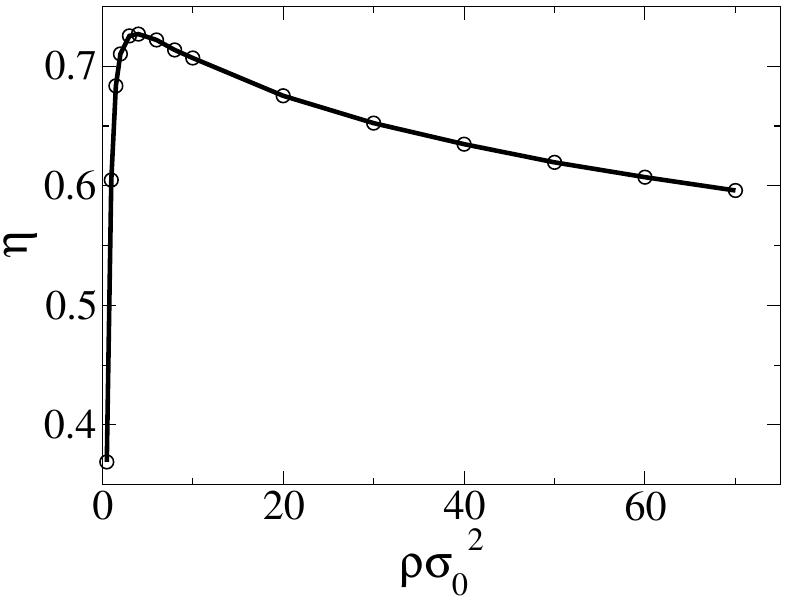} 
\caption{\label{fig3}
Packing fraction $\eta$ as a function of density $\rho \sigma_0^2$ for Gaussian-like polydisperse hard disks with $\nu/\sigma_0=0.1$.
}
\end{center}
\end{figure}

Moreover, as shown in Fig.~\ref{fig2}a, with increasing the density of the system, the particle size polydispersity in the system first decreases, which coincidences with the formation of the ordered hexatic phase indicated by the increase of $\langle \Psi_6 \rangle$, and decreases again when the system transforms into a disordered phase with the decrease of $\langle \Psi_6 \rangle$ at high density. 
{ This suggests that the decrease of the particle size polydispersity with increase density at low pressure is driven by the tendency of forming ordered hexatic phase in the system.}
The re-entrant phase behaviour of polydisperse systems was originally predicted in Ref.~\cite{bartlett1999}, in which a disordered glass was suggested at high density. To see whether the disordered phase is a kinetically arrested glass or diffusive fluid, we performed event driven molecular dynamics simulations starting from the equilibrated configurations from our ECMC simulations.
As shown in Fig.~\ref{fig2}b, during the re-entrant melting of hexatic phase at high density, the diffusion coefficient $D$ increases significantly, and the high density disordered phase is actually a diffusive fluid~\cite{SampedroRuiz2019}. To understand this intriguing re-entrant melting transition, we plot the packing fraction $\eta = \frac{\pi}{4} \rho \langle \sigma^2 \rangle $ as a function of density in Fig.~\ref{fig2}b, and one can see that during the re-entrant melting of hexatic phase, the packing fraction of the system decreases. More interestingly, even in the hexatic phase, the packing fraction $\eta$ does not increase monotonically with increasing density, and it reaches a maximum value at about $\rho \sigma_0^2 \simeq 2.7$. To check whether the non-monotonic dependence of $\eta$ on $\rho$ is a general feature for Gaussian-like polydisperse hard disks at large $\nu$, we perform simulations for Gaussian-like polydisperse hard disks with $\nu / \sigma_0 = 0.1$, in which, according to the phase diagram of the system (Fig.~\ref{fig:chapt4_1}), there is no phase transition with increasing $\rho$, and the obtained $\eta$ as a function of $\rho \sigma_0^2$ is shown in Fig.~\ref{fig3}. One can see that with increasing $\rho$, the packing fraction of the Gaussian-like polydisperse hard disks first increases and then decreases with a maximum at around $\rho \sigma_0^2 \simeq 3$. This suggests that the non-monotonic behaviour of packing fraction is indeed a general feature of highly polydisperse Gaussian-like hard disks, including the polydispersity range where the re-entrant melting transition occurs. This makes it unclear whether the re-entrant melting also exists in other systems of polydisperse hard disks. Because for modelling highly polydisperse hard-disk systems, the Gaussian-like particle size distribution may not be a good choice, as it produces an unphysical positive probability of having particle size at zero, which becomes more pronounced at large polydispersity.
To this end, in the following of this article, we investigate two other representative polydisperse systems, in which the probability of having $\sigma = 0$ is zero. 

\section{Lognormal polydisperse hard disks\label{sec2}}
\begin{figure*}
\begin{center}
\includegraphics[width=1.\linewidth]{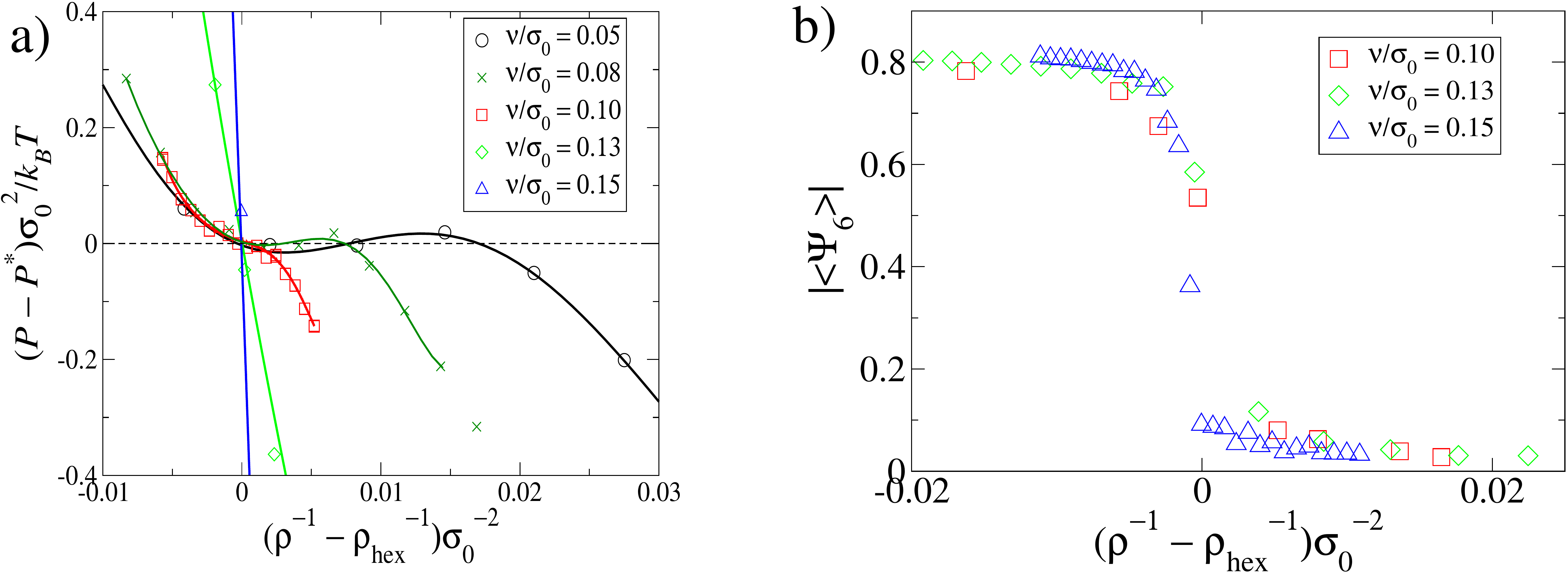} 
\caption{\label{fig4} 
(a) EOS for lognormal polydisperse hard-disk systems with various polydispersity parameter $\nu /\sigma_0 = 0.05$ to 0.15 in the representation of $(P - P^*) \sigma_0^2 / k_B T $ vs $(\rho^{-1} - \rho_{hex}^{-1})\sigma_0^{-2}$, where $P^*$ and $\rho_{hex}$ are the pressure and density of the hexatic phase at the fluid-hexatic transition, respectively, and the solid lines are fits of the EOS using 5th order polynomials. (b)  $|\langle \Psi_6\rangle|$  as functions of $(\rho^{-1} - \rho_{hex}^{-1})\sigma_0^{-2}$ for systems with $\nu /\sigma_0 = 0.10, 0.13$ and 0.15.}
\end{center}
\end{figure*}

\begin{figure}
\begin{center}
\includegraphics[width=1.\linewidth]{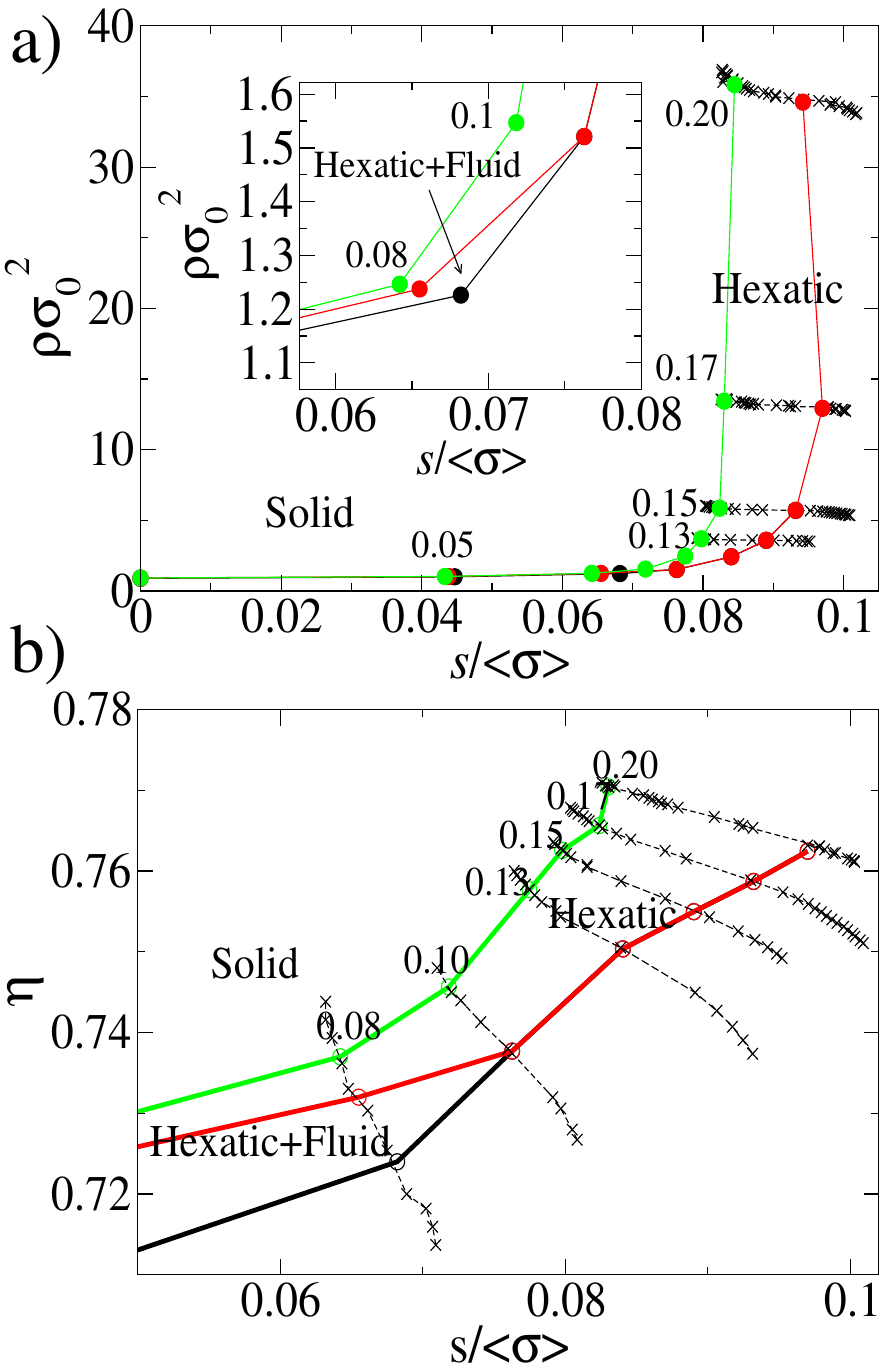} 
\caption{(a) Phase diagram of the polydisperse hard-disk systems of lognormal distribution in the representation of $\rho \sigma_0^2$ and $s/\langle \sigma \rangle$. The phase boundaries are obtained from $NVT-\Delta \mu$ simulations for systems with $\nu/ \sigma_0 = 0.005$ to 0.2. Inset: the enlarged view of the region of phase diagram at $0.06 \le s/\langle \sigma \rangle \le 0.08$. The crosses are the simulation results 
along the the re-entrant transitions at $\nu/\sigma_0 = 0.13$, 
0.15, 0.17, and 0.2 from bottom to top. (b) Low pressure phase diagram of polydisperse hard disks with lognormal distribution in the representation of $\eta$ vs $s/\langle \sigma \rangle$, where $\eta$ is the packing fraction of the system. The state points obtained from simulations at each $\nu/\sigma_0$ from 0.08 to 0.20 are shown as the symbols. The error bars are smaller than the symbols.
\label{fig4xxx}}
\end{center}
\end{figure}

\begin{figure}
\begin{center}
\includegraphics[width=1.\linewidth]{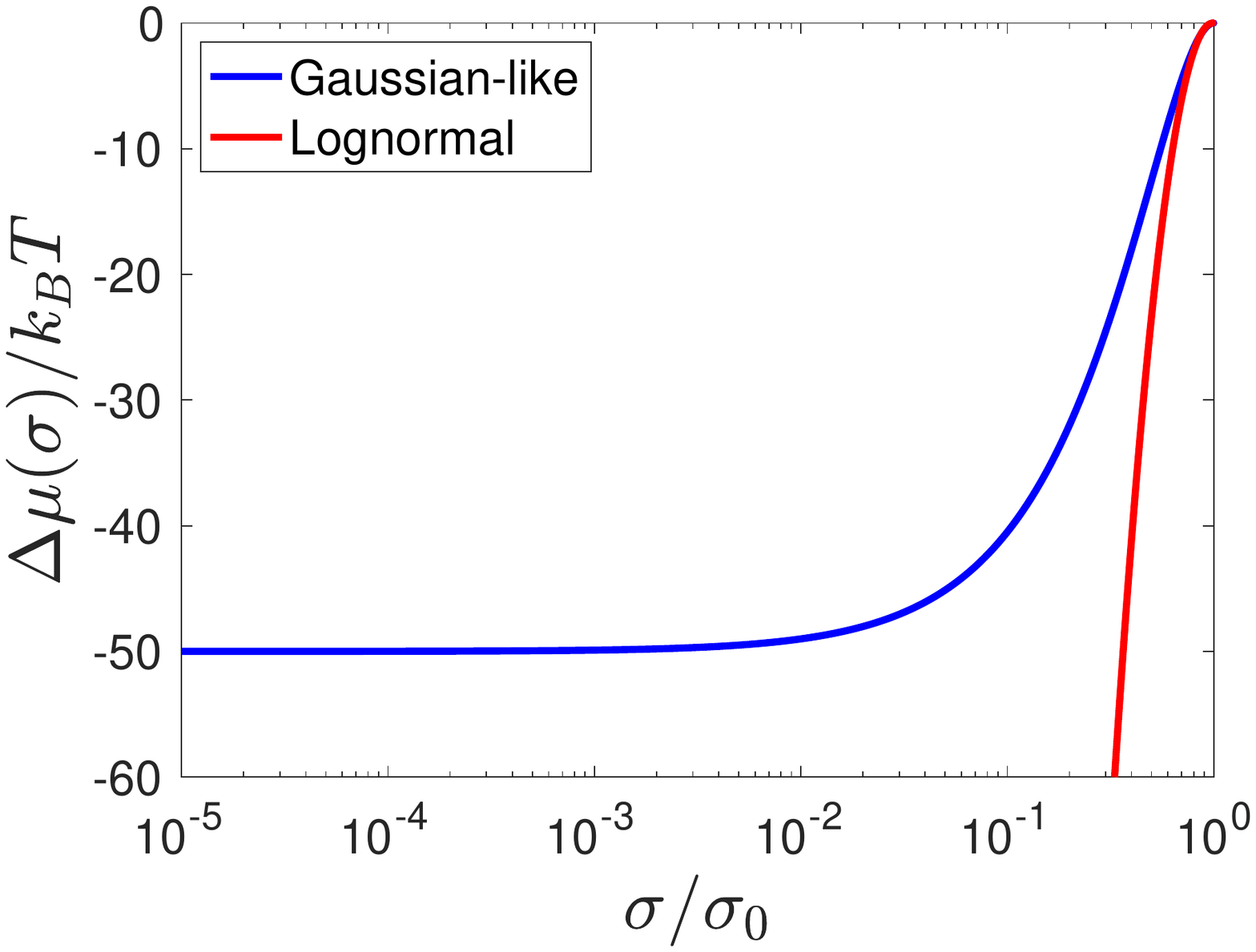} 
\caption{Chemical potential difference $\Delta \mu(\sigma)/k_BT$ for Gaussian-like (Eq.~\ref{eq:chapt4_1}) and lognormal polydisperse (Eq.~\ref{eq:chapt4_2}) hard disks with $\nu/\sigma_0 = 0.1$.\label{fig4xx}}
\end{center}
\end{figure}

The major artefact of Eq.~\ref{eq:chapt4_1} is that the probability of having $\sigma = 0$ is a finite positive number. To avoid this, 
we can use other probability distributions of polydisperse particles.
Here we investigate the phase behaviour of hard disks with two representative probability distributions. 

First, we introduce the lognormal distribution, in which $\log(\sigma / \sigma_0)$ obeys a Gaussian distribution. Therefore, the chemical potential between particles of different size is 
\begin{equation}
\frac{\Delta \mu (\sigma) }{k_BT}= -\frac{\log^2(\sigma/\sigma_0)}{2\nu^2}-\log \left(\frac{\sigma}{\sigma_0}\right)
\label{eq:chapt4_2}
\end{equation}
where $\nu$ is the polydispersity parameter, and in the ideal gas limit Eq.~\ref{eq:chapt4_2} gives a system of polydisperse particles centered around $\sigma = \sigma_0$ with the standard deviation $\nu$. 

We perform $NVT-\Delta \mu$ simulation using the ECMC algorithm with $N=256^2$ particles based on Eq.~\ref{eq:chapt4_2}, and the calculated EOS for systems of various polydispersity parameter $\nu$ is shown in Fig.~\ref{fig4}a. One can see that similar to the situation in Gaussian-like polydisperse hard disks, with increasing $\nu$, the Mayer-Wood loop in EOS becomes smaller implying that first-order transition becomes weaker~\cite{mayerloop}. For { $\nu / \sigma_0 \ge 0.1$}, the Mayer-Wood loop in EOS completely disappears, while there is clearly a transition from a disordered state to an ordered state indicated by the sharp increase of $|\langle \Psi_6\rangle|$ with increasing density as shown in Fig.~\ref{fig4}b. By checking the decay of $g(x,0)-1$ and $g_6(r)$, we ensure that the ordered phase forming from the fluid is the hexatic phase, which is qualitatively the same as in the system of Gaussian-like polydisperse hard disks~\cite{SampedroRuiz2019}.

The calculated phase diagram for polydisperse hard-disks systems of lognormal distribution is shown in Fig.~\ref{fig4xxx}a in the presentation of $\rho \sigma_0^2$ and $s/\langle \sigma \rangle$, { for which the scaling of $g(x,0)-1$ and $g_6(r)$ can be found in Table S1 in Supplementary Material}. One can see that qualitatively, the phase diagram is very similar to the system of Gaussian-like polydisperse hard disks in Fig.~\ref{fig:chapt4_1}, in which with increasing polydispersity, the first-order fluid-hexatic transition becomes weaker and eventually changes to be a continuous transition, and the density range for stable hexatic phase increases by orders of magnitude~\cite{SampedroRuiz2019}. Moreover, as shown in Fig.~\ref{fig4xxx}a, at $s/\langle \sigma \rangle \gtrsim 0.09$, with increasing $\rho$, the system can change from fluid to hexatic and to fluid again at very high density, which is seemingly qualitatively the same as in the system of Gaussian-like polydisperse hard disks (Fig.~\ref{fig:chapt4_1}). Moreover, we plot the low pressure phase diagram of the system in Fig.~\ref{fig4xxx}b, and one can see that with increasing $\nu/\sigma_0$, the packing fraction range for stable hexatic phase increases by one order of magnitude, which is similar to the Gaussian-like polydisperse hard disks~\cite{SampedroRuiz2019}. However, as shown in Fig.~\ref{fig4xxx}a, one can see that at fixed large $\nu/\sigma_0$, e.g. $\nu/\sigma_0 \gtrsim 0.13$, with increasing the density of the system the polydispersity $s/ \langle \sigma \rangle$ decreases, which is in contrast with the situation in the system of Gaussian-like polydisperse hard disks, where $s/ \langle \sigma \rangle$ increases with increasing density at large $\nu/\sigma_0$ as shown in Fig.~\ref{fig:chapt4_1}. This implies that in the sedimentation of highly polydisperse hard disks with lognormal distributions, no re-entrant melting of hexatic phase occurs. This can be understood from the chemical potential difference in the two different distributions. In Fig.~\ref{fig4xx}, we plot Eq.~\ref{eq:chapt4_1} and \ref{eq:chapt4_2} for $\nu/\sigma_0 = 0.1$. One can see that for Gaussian-like polydisperse system, when $\sigma$ is very small, e.g. $\sigma/\sigma_0 < 10^{-2}$, further decreasing $\sigma$ does not cost much free energy, which essentially drive the system to decrease the particle size at higher pressure to form a random fluid with increasing the true polydispersity of the system $s/\langle \sigma \rangle$.  On the contrary, for the polydisperse hard disks with lognormal distributions, the free energy cost for decreasing the the particle size increases dramatically and diverges when approaching $\sigma = 0$, this effectively prevents the decrease of particle size as well as the true polydispersity of the system $s/\langle \sigma \rangle$ at large pressure. Therefore, at very high pressure, polydisperse hard disks with lognormal distributions are almost ``monodisperse'' and forming an ordered solid phase.
However, according to the phase diagram in Fig.~\ref{fig4xxx}a, if one keeps the polydispersity of the system fixed at a certain value above $s/ \langle \sigma \rangle \simeq 0.09$ , increasing the density, or pressure, of the system, the system indeed can transform from a fluid phase to a hexatic phase and then to a fluid phase again at very high density or pressure.

\section{Triangular polydisperse hard disks \label{sec3}}
\begin{figure}
\begin{center}
\includegraphics[width=1.\linewidth]{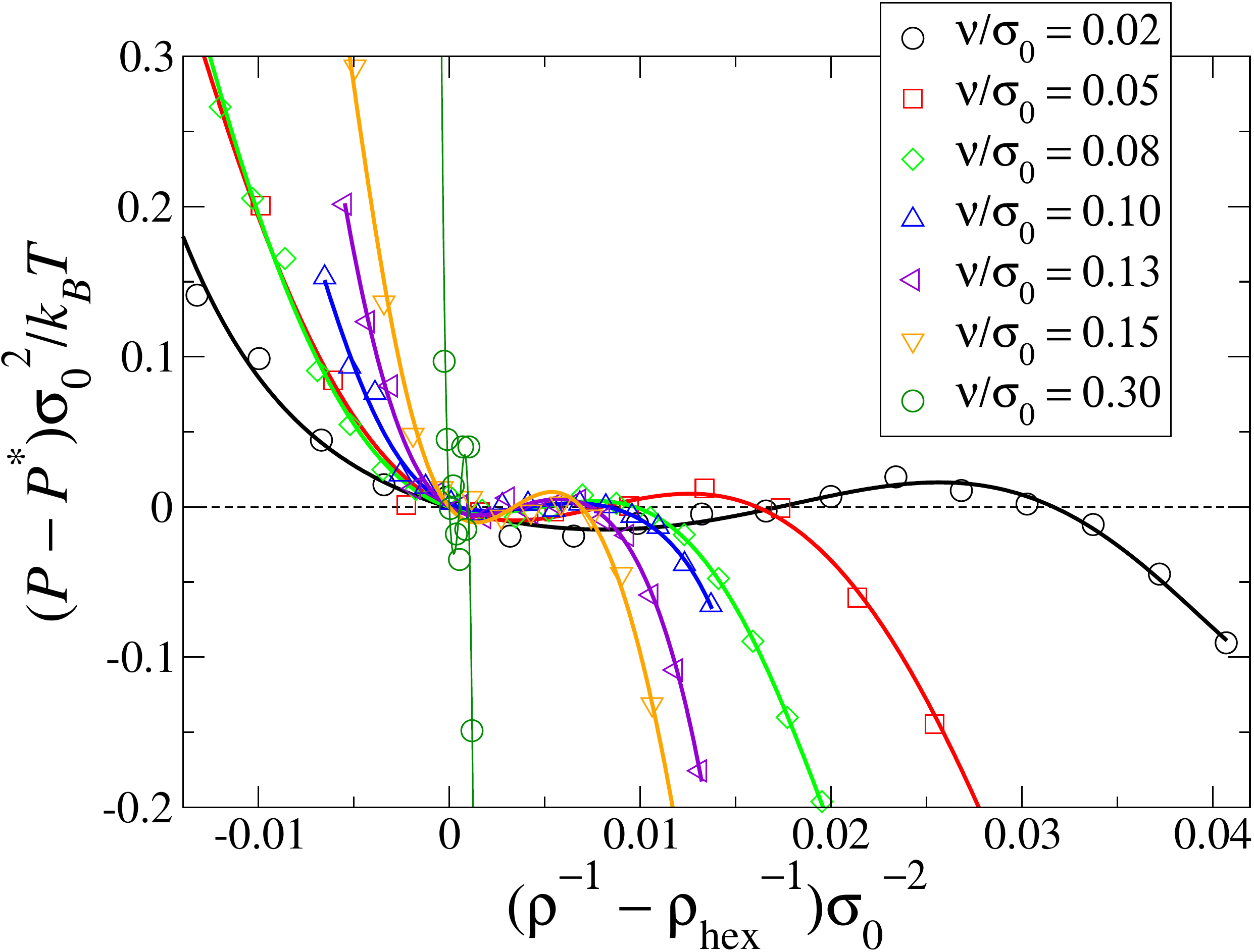} 
\caption{\label{figtriang}
EOS for triangle polydisperse hard-disk systems with various polydispersity parameter $\nu /\sigma_0 = 0.25$ to 0.3 in the representation of $(P - P^*) \sigma_0^2 / k_B T $ vs $(\rho^{-1} - \rho_{hex}^{-1})\sigma_0^{-2}$, where $P^*$ and $\rho_{hex}$ are the pressure and density of the hexatic phase at the fluid-hexatic transition, respectively, and the solid lines are fits of the EOS using 5th order polynomials. 
}
\end{center}
\end{figure}

\begin{figure}
\begin{center}
\includegraphics[width=1.\linewidth]{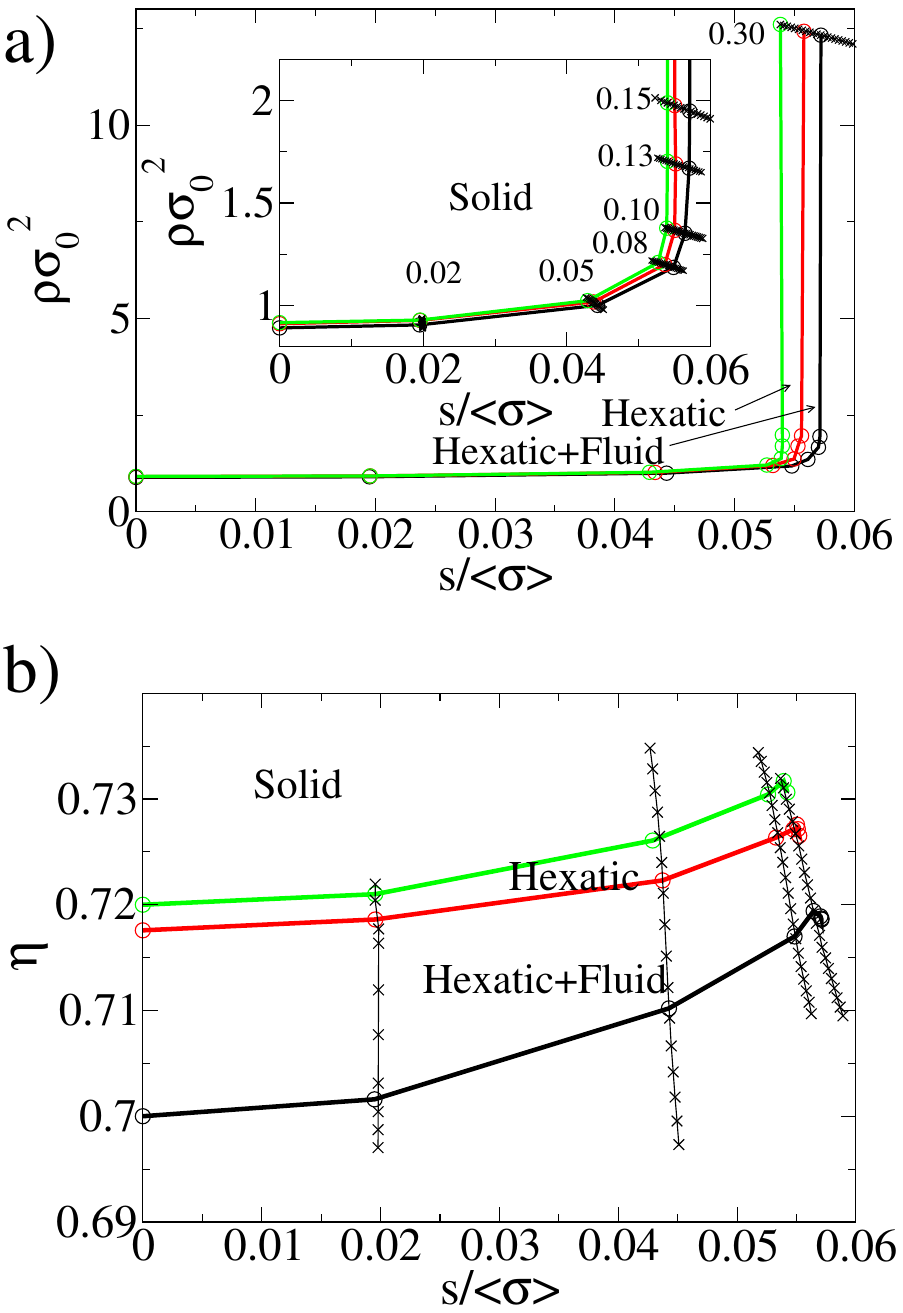} 
\caption{Phase diagram for the systems of triangular polydisperse hard disks in the representations of $\rho\sigma_0^2$ vs $s/\langle \sigma \rangle$ (a) and $\eta$ vs $s/\langle \sigma \rangle$ for $\nu/\sigma_0$ from 0.02 to 0.3. The inset in (a) is the enlarged view of the low polydispersity region. \label{fig7}}
\end{center}
\end{figure}

\begin{figure}
\begin{center}
\includegraphics[width=1.\linewidth]{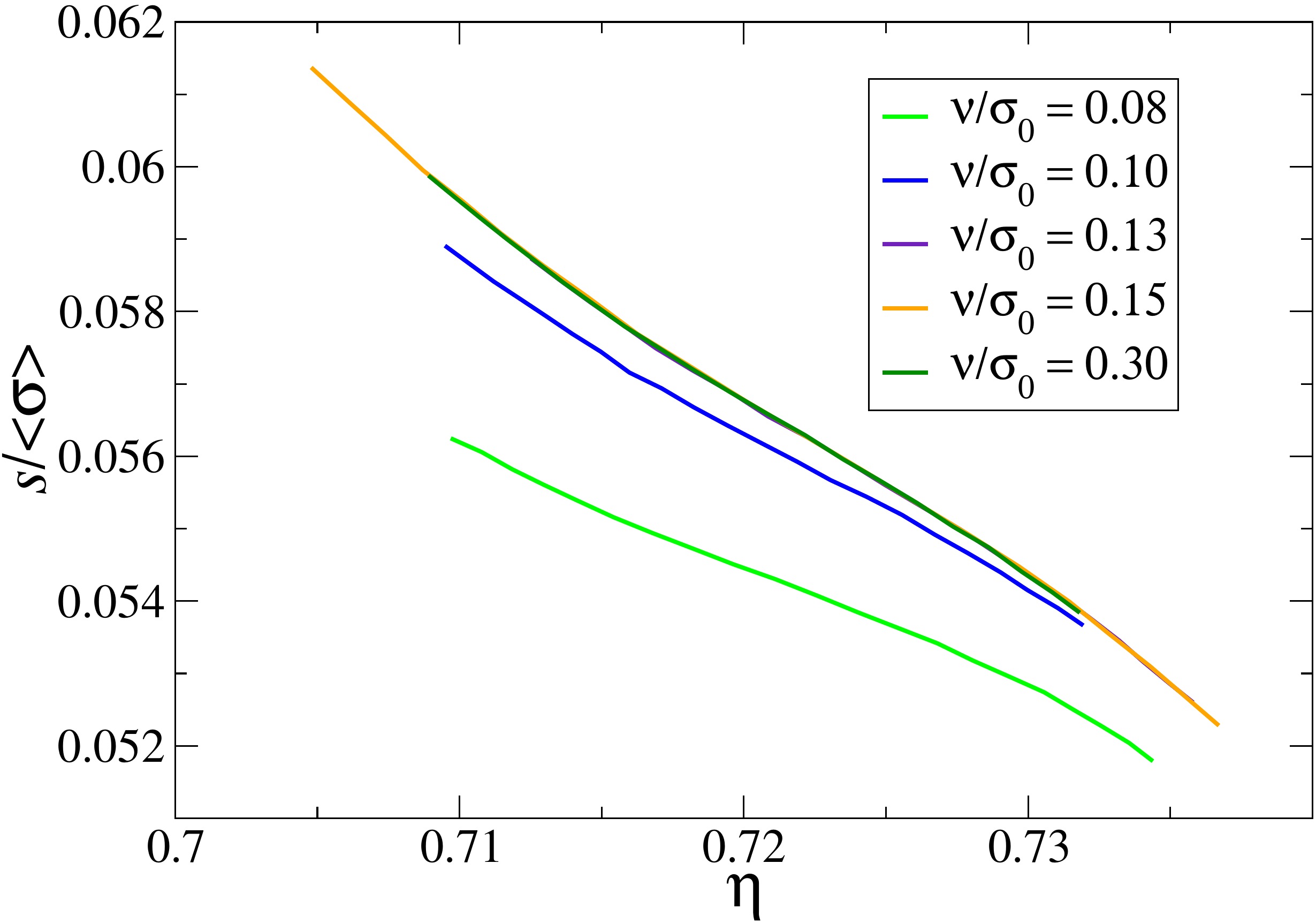} 
  \caption{Polydispersity $s/\langle \sigma \rangle $ as a function of packing fraction $\eta$ for highly polydisperse hard disks with triangle distributions at $0.08 \le \nu /\sigma \le 0.3$.\label{fig8}}
\end{center}
\end{figure}

In Sec.~\ref{sec2}, we studied the phase behavior of polydisperse hard disks with lognormal distributions, where the particle size probability vanishes at $\sigma = 0$. Here we investigate another different type of polydisperse system, i.e., triangle distribution, in which the particle size probability with a finite range above $\sigma = 0$ is zero. The chemical potential between particles of different size is 
\begin{equation}\label{eqtriangle}
\frac{\Delta \mu (\sigma)}{k_BT} = \left\{
\begin{array}{c c c}
-\ln \left(\frac{\sqrt{6}\nu-|\sigma-\sigma_0|}{\sqrt{6}\nu }\right) & \text{if} & |\sigma-\sigma_0|< \sqrt{6}\nu, \\
-\infty & \text{else},\\
\end{array}
\right.
\end{equation}
where $\nu$ is the polydispersity parameter. This distribution is characterized for having a maximum at $\sigma =\sigma_0$ and linearly decays as it moves away from the maximum until reaching zero probability for a distance $\sqrt{6}\nu$. In the ideal gas limit Eq.~\ref{eqtriangle} gives a system of polydisperse particles centered around $\sigma = \sigma_0$ with the standard deviation $\nu$.

First, we perform $NVT-\Delta \mu$ ECMC simulations in a $256^2$ particle system to obtain the equation of state for values of $\nu$ ranging between $0.01$ and $0.30\sigma_0$ as shown in Fig.~\ref{figtriang}. 
One can see that with increasing $\nu$, the first-order transition from the fluid phase does not become weaker, and even at $\nu/ \sigma_0 = 0.3$, there is still a pronounced Mayer-Wood loop in the EOS.
By checking the decay of $g(x,0)-1$ and $g_6(r)$, we ensure that the ordered phase forming from the fluid is a hexatic phase, and this persistent first-order fluid-hexatic phase transition in triangle polydisperse hard disks is markedly different from the polydisperse hard-disk systems of Gaussian-like and lognormal polydisperse distributions.

The phase diagrams of polydisperse hard disks with various  triangle distributions are summarized in Fig.~\ref{fig7}, { for which the scaling of $g(x,0) - 1$ and $g_6(r)$ can be found in Table S2 in the Supplementary Material}. One can see that the density range for stable hexatic phase does not change significantly with increasing $\nu/\sigma_0 = 0$ to 0.08, and the corresponding packing fraction range $\eta$ for stable hexatic phase increases from 0.002 at $\nu/\sigma_0 = 0$ to about 0.005 at $\nu / \sigma_0 = 0.3$ (Fig.~\ref{fig7}b).  
Moreover, the re-entrant melting transition of hexatic phase does not exist in the range of $\nu/\sigma_0$ studied from $\nu/\sigma_0 = 0.02$ to 0.3. At high polydispersity, e.g. $\nu/\sigma_0 \gtrsim 0.13$, as shown in Fig.~\ref{fig7}b, the phase boundary between fluid and hexatic phase converges to ($\eta_{fluid} \simeq 0.72$, $s_{fluid}/ \langle \sigma \rangle \simeq 0.057$) and ($\eta_{hex} \simeq 0.728$, $s_{hex}/ \langle \sigma \rangle \simeq 0.055$), and hexatic to solid transition point converges to ($\eta_{solid} \simeq 0.731$, $s_{solid}/ \langle \sigma \rangle \simeq 0.054$). In Fig.~\ref{fig8}, we plot the true polydispersity of the system $s/\langle \sigma \rangle$ as a function of packing fraction $\eta$ for triangle polydisperse hard disks with various $\nu/\sigma_0$, and one can see that for $\nu/\sigma_0 \ge 0.1$, all curves collapse into a single curve, suggesting that at high enough $\nu/\sigma_0$, the property of system essentially does not depend on the polydispersity parameter $\nu$. To understand this, we define a normalized particle size 
\begin{equation}
\sigma^* = \frac{\sigma}{\sigma_{\min}},
\end{equation}
where $\sigma_{\min} = \sigma_0 - \sqrt{6} \nu$ is the minimal particle size for the triangle distribution of $\nu$. Then we can re-write Eq.~\ref{eqtriangle} into
\begin{equation}\label{eqtriangle2}
\frac{\Delta \mu (\sigma)}{k_BT} = \left\{
\begin{array}{c c c}
-\ln \left(\frac{\sqrt{6}\nu-|\sigma_0 - \sigma^* \sigma_{\min}|}{\sqrt{6}\nu }\right) & \text{if} & |\sigma^* - \frac{\sigma_0}{\sigma_{\min}}| < \frac{\sqrt{6}\nu}{\sigma_{\min}},\\
-\infty & \text{else}.\\
\end{array}
\right.
\end{equation}
At high pressure, when $\sigma$ is approaching $\sigma_{\min}$, i.e., $ \sigma^*  \simeq 1$, we have
\begin{equation}
\frac{d \Delta \mu}{d \sigma^*} = \frac{k_BT}{1-\sigma^*},
\end{equation}
which implies that at high pressure, the change of $\Delta \mu$ does not depend on polydispersity parameter $\nu$. This suggests at high enough pressure, when the average particle size is close to $\sigma_{\min}$, i.e.,  $\langle \sigma^* \rangle \simeq 1$, the systems for different $\nu$ are essentially the same. As in systems of highly polydisperse hard disks, the distribution of particle size is wider, the pressure or packing fraction required to push $\langle \sigma^* \rangle$ towards 1 is smaller, and this explains the collapse of curves for different $\nu$ in Fig.~\ref{fig8} at high polydispersity and high packing fraction.

%

\section{Conclusions\label{sec4}}
Following our previous work on the phase transitions in Gaussian-like polydisperse hard disks~\cite{SampedroRuiz2019}, in this article,
we have investigated the effect of particle size distribution on the phase behaviour of polydisperse hard disks.
The main results of Ref.~\cite{SampedroRuiz2019} are: (i) with increasing the polydispersity of the system, the {first order } hexatic-fluid transition becomes weaker and completely switches to a continuous transition following the celebrated KTHNY scenario at high polydispersity; (ii) simultaneously, the stable density range of hexatic phase is enlarged by orders of magnitude; (iii) at high polydispersity parameter $\nu$ re-entrant melting transitions of ordered phases, i.e., solid and hexatic phases, are observed with increasing pressure, which suggests that the re-entrant melting transition can be found in sedimentation experiments~\cite{dullens2017}.
However, in the Gaussian-like polydisperse distribution, the probability of having the particle size of zero is finite, which is the artefact of the model. Therefore, this motivates us to investigate the phase behaviour of polydisperse hard disks with two other representative types of particle size distributions with the vanishing probability at the particle size of zero in this work, i.e., lognormal and triangle distributions.

In systems of polydisperse hard disks with lognormal distributions, the phase diagram appears qualitatively the same as the Gaussian-like polydisperse hard disks, and the enhanced stability of hexatic as well as the switch of hexatic-fluid transition {from first order to continuous} are both found in the system of polydisperse hard disks with lognormal distribution with increasing polydispersity. Moreover, at a fixed true polydispersity of the system, $s/\langle \sigma \rangle$, with increasing density, re-entrant melting of solid and hexatic phases can be also found in highly polydisperse lognormal systems of hard disks. However, at any polydispersity parameter $\nu$, with increasing pressure or density, the true polydispersity of the system monotonically decreases, and no re-entrant melting transition is found.
The reason is that at very high pressure, the free energy cost of decreasing the particle size is diverging at zero for lognormal distributions, while it is almost zero for Gaussian-like distributions.
This suggests that different from the Gaussian-like polydisperse hard disks, in the sedimentation of polydisperse hard disks of lognormal distributions, we would not find any re-entrant melting transitions of ordered phases.

Moreover, in systems of polydisperse hard disks with triangle distributions, the phase diagram is qualitatively different from those of Gaussian-like and lognormal distributions. For all polydispersity parameter studied, the hexatic-fluid transition in polydisperse hard disks with triangle distributions is always strongly first order, and we do not observe significant increase of stable density range of hexatic phase with increasing polydispersity of the system.
Moreover, we can not reach any system of true polydispersity higher than 0.06. The reason is due to the fact that at high pressure and high polydispersity, the change of chemical potential of different particle size does not depend on the polydispersity of the system.
{In addition, in polydisperse hard-disk systems of triangle distributions, we also do not find any re-entrant behaviour with increasing pressure.
Therefore, we believe that the non-zero probability of having particle size zero is the reason causing the re-entrant melting transition in the sedimentation of Gaussian-like polydisperse hard disks, which is an artefact of the Gaussian-like polydisperse hard-disk model.} 
Our results show that the exact particle size distribution plays an important role in the phase behaviour of polydisperse hard disks, and 
so far we can not reach a universal phase diagram for all polydisperse hard disks.

\section{Supplementary Material}
See the Supplementary Material for the $g(x,0)-1$ and $g_6(r)$ for obtaing the phase diagrams of Fig. 5 and Fig. 8.

\begin{acknowledgements}
This work has been supported in part by the Singapore Ministry of Education through the Academic Research Fund MOE2019-T2-2-010 and RG104/17 (S), by Nanyang Technological University Start-Up Grant (NTU-SUG: M4081781.120), by the Advanced Manufacturing and Engineering Young Individual Research Grant (A1784C0018) and by the Science and Engineering Research Council of Agency for Science, Technology and Research Singapore. We thank NSCC for granting computational resources.
\end{acknowledgements}

\section{Data and materials availability}
All data needed to evaluate the conclusions in the paper are presented in the paper and/or the Supplementary Material. Additional data related to this paper may be requested from the authors.

\bibliographystyle{h-physrev}
\bibliography{ref}

\end{document}